Superconductivity in 122-type antimonide $BaPt_2Sb_2$

Motoharu Imai[1], Soshi Ibuka[1*], Naoki Kikugawa[1], Taichi Terashima[2], Shinya Uji[2], Takeshi Yajima[3, 4], Hiroshi Kageyama[3], and Izumi Hase[5]

[1] Superconducting Property Unit, National Institute for Materials Science, 1-2-1 Sengen, Tsukuba, Ibaraki 305-0048, Japan.

[2] Superconducting Property Unit, National Institute for Materials Science, 3-13 Sakura, Tsukuba, Ibaraki 305-0003, Japan.

[3]Graduate School of Engineering, Kyoto University, Nishikyo-ku, Kyoto 615-8510, Japan

[4]Institute for Solid State Physics, University of Tokyo, 5-1-5 Kashiwanoha, Kashiwa, Chiba 277-85851, Japan.

[5]Electronics and Photonics Research Institute, National Institute of Advanced Industrial Science and Technology, AIST Central 2, 1-1-4 Umezono, Tsukuba, Ibaraki 305-8568, Japan.

* Present address: Institute of Materials Structure Science, High Energy Accelerator Research Organization, 203-1 Shirakata, Tokai, Ibaraki 319-1106, Japan

The crystal structure, superconducting properties, and electronic structure of a novel superconducting 122-type antimonide, $BaPt_2Sb_2$, have been investigated by measurements of powder X-ray diffraction patterns, electrical resistivity, ac magnetic susceptibility, specific heat as well as ab-initio calculations. This material crystallizes in a new-type of monoclinic variant of the $CaBe_2Ge_2$-type structure, in which $Pt_2Sb_2$ layers consisting of $PtSb_4$ tetrahedra and $Sb_2Pt_2$ layers consisting of $SbPt_4$ tetrahedra are stacked alternatively and Ba atoms are located between the layers. Measurements of electrical resistivity, ac magnetic susceptibility and specific heat revealed that $BaPt_2Sb_2$ is a superconducting material with a $T_C$ of 1.8 K. The electronic heat capacity coefficient $\gamma_n$ and Debye temperature $\theta_D$ were 8.6(2) mJ/mol $K^2$ and 146(4) K, where the figures in parentheses represent the standard deviation. The upper critical field $\mu_0 H_{C2}(0)$ and the Ginzburg-Landau coherent length $\xi(0)$ were determined to be 0.27 T and 35 nm. Calculations showed that it has two three-dimensional Fermi surfaces (FSs) and two two-dimensional FSs, leading to anisotropic transport properties. The d-states of the Pt atoms in the $Pt_2Sb_2$ layers mainly contribute to $N(E_F)$. A comparison between experimental and calculated results indicates that $BaPt_2Sb_2$ is a superconducting material with moderate coupling.

I. INTRODUCTION

The discovery of superconductivity with a critical temperature $T_C$ of 26 K in LaFeAs(O,F)[1] stimulated studies on superconducting iron pnictides. Consequently, various types of superconducting iron pnictides, such as 1111, 122, 11 systems have been discovered.[2-9] $AFe_2As_2$ (A = Ba, Sr, Ca, or Eu) is the parent material of superconducting 122 iron pnictides. Partial substitution of A, Fe, or As or pressurization induce superconductivity with a relatively high $T_C$. For example, $(Ba,K)Fe_2As_2$ and $(Ca, La)Fe_2(As, P)_2$ have $T_C$ of 38 K and 45 K, respectively.[10, 11] $AFe_2As_2$ crystallizes in the $ThCr_2Si_2$-type structure, in which A atoms are located between $Fe_2As_2$ layers. The $Fe_2As_2$ layer is composed of an Fe square lattice, where each Fe atom is surrounded by four As atoms in tetrahedral coordination.

Superconductivity has also been observed in iron-free pnictides with the $ThCr_2Si_2$-type or related structure, although their $T_C$ values are relatively low.[12-19] $SrPt_2As_2$ is a superconductor with a $T_C$ of 5.4 K.[20] It crystallizes in a modulated orthorhombic variant of the $CaBe_2Ge_2$-type structure, a derivative structure of the $ThCr_2Si_2$-type.[21] In $SrPt_2As_2$ with the $CaBe_2Ge_2$-type structure (referred to hereafter as $CaBe_2Ge_2$-type $SrPt_2As_2$), Sr atoms are sandwiched by two kinds of layers. One is a $Pt_2As_2$ layer, which is the same as that in $AFe_2As_2$. The other is an $As_2Pt_2$ layer, in which As atoms form a square lattice and each As atom is surrounded by four Pt atoms in tetrahedral coordination. The electronic structure has been calculated for $ThCr_2Si_2$-type $SrPt_2As_2$ ($SrPt_2As_2$ with the $ThCr_2Si_2$-type structure), which is a hypothetical material, as well as $CaBe_2Ge_2$-type $SrPt_2As_2$.[22] The calculations show that the Fermi surfaces (FSs) of the $CaBe_2Ge_2$-type $SrPt_2As_2$ have intermediate character between $AFe_2As_2$ and $ThCr_2Si_2$-type iron-free pnictides. $AFe_2As_2$ has two-dimensional (2D) FSs[2,3,5,6,23,24] while $ThCr_2Si_2$-type iron-free pnictides have three-dimensional (3D) FSs.[25-29] The $CaBe_2Ge_2$-type $SrPt_2As_2$ has two 2D-like FSs and two 3D-like ones. The coexistence of 2D and 3D FS has also been reported in $CaBe_2Ge_2$-type $BaPd_2Sb_2$.[29] Thus, some of the $CaBe_2Ge_2$-type iron-free pnictides are positioned between the $ThCr_2Si_2$-type iron pnictides and the $ThCr_2Si_2$-type iron-free pnictides from the view-point of the Fermi surfaceology. Therefore, the study of superconductivity in the $CaBe_2Ge_2$-type pnictides is significant for a better and systematic understanding of superconductivity in 122 pnictides.

Superconducting materials structurally related to the iron-based superconducting materials have thus far been limited to phosphides and arsenides. Recently, a 122 antimonide, $SrPt_2Sb_2$, has been reported to be a superconductor with a $T_C$ of 2.1 K.[30] This is the first reported superconducting 122 antimonide that is related to superconducting iron pnictides, to the best of our knowledge. The crystal structure of $SrPt_2Sb_2$ is different from that of $CaBe_2Ge_2$-type, which was reported previously.[21] Although a precise analysis of the crystal structure needs to be carried out, it is plausibly related to the $CaBe_2Ge_2$-type structure. Since $SrPt_2Sb_2$ is isovalent to $SrPt_2As_2$ and has a crystal structure related to the $CaBe_2Ge_2$-type, $SrPt_2Sb_2$ is expected to have both 3D and 2D FSs and to show physical properties that reflects this.

In this paper, we report on the successful synthesis of $BaPt_2Sb_2$, its crystal and electronic structures, and the appearance of superconductivity in the material. $BaPt_2Sb_2$ is a novel material to the best of our knowledge. Powder X-ray diffractometry indicates that it crystallizes in a monoclinic variant of the $CaBe_2Ge_2$-type structure. Measurements of electrical resistivity, ac magnetic susceptibility, and specific heat reveal that $BaPt_2Sb_2$ is a superconducting material with a $T_C$ of 1.82 K. Ab-initio calculations demonstrate that it has 3D and 2D FSs and the Pt-d states largely contribute to $N(E_F)$, which is responsible for its superconductivity.

## II. EXPERIMENTAL METHODS

The starting materials used are Ba (99.9% purity), Pt (99.95% purity), and Sb (99.999% purity). Samples were synthesized by Ar-arc melting of a 1.05:2:2 molar mixture of Ba, Pt, and Sb. Electron micro-probe analysis (EPMA) revealed that the sample consisted of a matrix with chemical composition of 20.2 at.% Ba, 40.6 at.% Pt, and 39.3 at.% Sb, and a small amount of precipitates of $Pt_xSb$.

The crystal structure was examined by powder X-ray diffractometry. The X-ray diffraction (XRD) measurements were performed on a Bragg-Brentano diffractometer RINT-TTR III (Rigaku) using Cu K$\alpha$ radiation (40 kV, 300 mA) with a step width of 0.02° in the 2θ range of 15.0–120.0° at room temperature. For the measurements, a fine powder sample was mounted on a glass plate holder of 0.3 mm depth. The observed XRD pattern was analyzed by the direct-space method with simulated annealing algorithm and subsequently refined by the Rietveld method[31] using the software "FullProf".[32] The background was modeled as a six-coefficient polynomial function.

The electrical resistivity ρ at temperatures ranging from 1.8 to 300.0 K was measured by a four-probe method using a physical properties measurement system (PPMS, Quantum Design Co.). The magnetic field dependence of ρ was measured using a standard ac technique with a frequency ($f$) of ~15 Hz.

We measured the ac susceptibility $\chi_{ac} = \chi' - i\chi''$ by a mutual-inductance method with an ac modulation field of ($\mu_0 H_{ac}$) ~ 0.037 mT and $f$ ~ 67 Hz. For this, we used the same sample that was used for the ac resistivity measurements. The sample was cooled down with a dilution fridge that was equipped with a superconducting magnet to apply the external fields. The specific heat was measured by a relaxation method using PPMS.

In order to investigate the electronic structure of $BaPt_2Sb_2$, we performed an *ab-initio* band calculation for this compound. We used the full-potential augmented plane-wave (FLAPW) scheme and the exchange-correlation potential was constructed within the local-density approximation (LDA). These are implemented as KANSAI-94 and TSPACE[33] computer codes. The space group is C2/m (No. 12), and the lattice constants and atomic parameters were fixed at the experimentally observed ones obtained in this work. Muffin-tin radii used in this calculation are 2.6 bohr for Ba, 2.4 bohr for Pt and 2.3 bohr for Sb, respectively. For the plane wave basis functions, we used about 1000 LAPWs. Since this compound contains the heavy atoms Pt and Sb, we included the spin-orbit interaction (SOI) in the second-variational approach.[34] In this scheme, we first perform the scalar-relativistic calculation, and then include SOI iteratively.

## III. RESULTS AND DISCUSSION

### A. Crystal Structure

Figure 1 shows a powder XRD pattern of $BaPt_2Sb_2$ together with the results of Rietveld analysis. The observed pattern can be reproduced well assuming that $BaPt_2Sb_2$ has a monoclinic structure with the parameters shown in Table I and that a small amount (0.8 mol %) of impurity phase PtSb is mixed. Figure 2 illustrates the crystal structure of $BaPt_2Sb_2$. The blue solid lines represent a unit cell. The crystal structure of $BaPt_2Sb_2$ is a new monoclinic variant of the $CaBe_2Ge_2$-type. In this structure, the Pt1 and Sb2 atoms form the $Pt_2Sb_2$ layers that consist of $PtSb_4$ tetrahedra, while the Pt2 and Sb1 atoms form the $Sb_2Pt_2$ layers that consist of $SbPt_4$ tetrahedra, and these layers are stacked alternately along the [001] direction with Ba atoms located between the layers. These feature is the same as those of the $CaBe_2Ge_2$-type. The following features are different from the $CaBe_2Ge_2$-type. First, the angle between the *c*-axis and the *a*-*b* plane, $\beta_{uc}$, is 91.227°, while the three angles $\alpha_{uc}$, $\beta_{uc}$, and $\gamma_{uc}$, are 90.00° in the $CaBe_2Ge_2$-type (Fig. 2 (d)). Second, the Pt1 (or Sb1) atoms form a deformed square lattice in the $Pt_2Sb_2$ (or $Sb_2Pt_2$) layers in the proposed structure while Pt (or Sb) atoms form a square lattice in the $Pt_2Sb_2$ (or $Sb_2Pt_2$) layers in the $CaBe_2Ge_2$-type $BaPt_2Sb_2$ (hypothetical) as shown in Figs. 2(e) and 2(f). The shape of the lattice formed by Pt1 (or Sb1) atoms is an isosceles trapezoid, and the degree of deformation in the Pt1 lattice is larger than that of the Sb1 lattice. Pt1 (or Sb1) atoms are arranged in a line in the *b*-axis direction while they are arranged in a zigzag fashion in the *a*-axis direction. Third, the Pt1 (or Sb1) deformed square lattice in the proposed structure is almost parallel to the *a*- and *b*-axis while the square lattice in the $CaBe_2Ge_2$-type rotates 45° with respect to the *a*- and *b*-axis (Figs. 2(e) and 2(f), respectively).

A monoclinic variant of the $CaBe_2Ge_2$-type has been reported in materials such as $LnPt_2Ge_2$ (Ln = La-Dy),[35] and a high-pressure phase of $SrPt_2As_2$,[21] which is known as the $LaPt_2Ge_2$-type structure. The $LaPt_2Ge_2$-type structure can be obtained from the $CaBe_2Ge_2$-type by changing the *b*-axis length and increasing the unit-cell angle β from 90° by several tenths of a degree. The relation between the unit cell axis and the square Pt lattice is different between the $LaPt_2Ge_2$-type and the proposed crystal structure of $BaPt_2Sb_2$; in the $LaPt_2Ge_2$-type, the square lattice rotates 45° with respect to the *a* and *b* axis as in the case of the $CaBe_2Ge_2$-type, while the square lattice is almost parallel to the *a* and *b* axis in $BaPt_2Sb_2$. Thus, this proposed structure is a novel one to the best of our knowledge.

Table II lists the interatomic distances. The intralayer interatomic distance has two values, 2.681(3) and 2.734(3) Å. The interlayer interatomic distance (2.708(7) Å) is almost the same as the intralayer interatomic distances. In this meaning, the crystal structure is three dimensional. The $T_C$ of iron-based superconductors has often been discussed with two structural parameters, the Pn-Fe-Pn angle in the $FePn_4$ tetrahedron and the Pn height from the Fe square lattice, where Pn represents pnictogen.[36-38] The Sb-Pt-Sb angles in the $PtSb_4$ tetrahedron and the Pt-Sb-Pt angles in the $SbPt_4$ tetrahedron are far from the ideal regular tetrahedron angle of 109.5° that is thought to be favorable for high $T_C$ in the iron pnictides. The Sb height from the distorted Pt square lattice is in the range of the Pn height of the iron pnictides with high $T_C$.

The crystal structure of $SrPt_2Sb_2$ is reported to be different from the $CaBe_2Ge_2$-type,[30] contrary to the claim by Imre et al.,[21] and it has not been determined yet. The structure proposed in the present work may be helpful to clarify the crystal structure of $SrPt_2Sb_2$.

B. Electrical Resistivity, ac magnetic susceptibility, specific heat, and magnetic phase diagram

We found that $BaPt_2Sb_2$ undergoes a superconducting transition based on the results of the measurements of electrical resistivity, ac magnetic susceptibility, and specific heat.

Figure 3(a) shows the electrical resistivity $\rho$ as a function of temperature, and the inset shows an extended view at low temperatures. $\rho$ decreases with decreasing temperature; it shows an abrupt decrease at 1.85 K and it becomes negligibly small at 1.82 K, which suggests that $BaPt_2Sb_2$ undergoes a normal-to-superconducting transition. The residual resistivity ratio RRR [$\rho$(300 K)/$\rho$(2 K)] is 4.04. Figure 3(b) shows $\rho$ measured at constant temperatures as a function of the magnetic field. When a magnetic field was applied, the sample underwent a superconducting-to-normal transition. The symbols $H_{C2}^{\rho_on}$, $H_{C2}^{\rho_mid}$, $H_{C2}^{\rho_comp}$ represent the magnetic field at which $\rho$ starts to increase from zero resistivity, the value of $\rho$ at half of the normal resistivity value, and an extrapolation of the $\rho$-$T$ curve at the transition reaching the normal resistivity value, respectively. The critical field increases with decreasing temperature. The details are discussed later.

Figure 4 shows the real and imaginary part of the ac magnetic susceptibility, $\chi$' and $\chi$'', measured at constant temperatures below 1.44 K as a function of magnetic field amplitude $\mu_0 H$. The inset shows an expanded view of $\chi$' measured at 0.03K. $\chi$' shows a diamagnetic signal at low $\mu_0 H$ and the diamagnetic signal disappears at $\mu_0 H_{C2}^{\chi'}$ with increasing $\mu_0 H$. $\chi$'' shows a peak at the magnetic field amplitude nearly equal to the inflection point of the $\mu_0 H$ - $\chi$' curve. The shoulder observed at low $\mu_0 H$ in the $\mu_0 H$ - $\chi$'' curve is a background signal that comes from the apparatus. This behavior of $\chi$' and $\chi$'' suggests that $BaPt_2Sb_2$ undergoes a superconducting transition in this temperature range,[39] and that the abrupt decrease of $\rho$ at 1.85 K corresponds to a superconducting transition. The magnitude of the diamagnetic signal of $BaPt_2Sb_2$ in $\chi$' was investigated by comparing it with the diamagnetic signal of Pb with the same shape and size. Since the magnitude of the observed diamagnetic signal of $BaPt_2Sb_2$ is almost the same as that of Pb measured at 1.6 K, at which the Pb sample is a perfect diamagnet, the volume fraction of superconductivity is approximately 100 % in $BaPt_2Sb_2$.

Figure 5(a) shows the specific heat divided by the temperature, $C/T$, measured at various magnetic fields as a function of the temperature squared, $T^2$. $C/T$ has an anomaly around 1.7 K for 0 T. This anomaly shifts to a lower temperature and becomes smaller with increasing magnetic field strength and it disappears at 0.2 T. The large specific heat anomaly and the large diamagnetic signal in $\chi$' indicate that the observed superconductivity is bulk superconductivity. These results reveal that $BaPt_2Sb_2$ is a superconducting material with $T_C$ of 1.8 K.

We determined the electronic specific heat coefficient $\gamma_n$, the coefficients $\beta$ and $\delta$ to be 8.6(2) mJ/mol $K^2$, 2.32(6) mJ/mol $K^4$, and 0.111(5) mJ/mol $K^6$, respectively, by fitting the data obtained at 0.2 T into the equation $C/T = \gamma_n + \beta T^2 + \delta T^4$, where the figures in parentheses represent the standard deviation. The Debye temperature $\theta_D$ was calculated to be 146(4) K, from the value of β. In order to confirm the validity of the fitting, we calculated the difference in entropy between the normal and superconducting states, $S_{es}$-$S_{en}$, by integrating $(C_{es}-\gamma_n)/T$ between 0 and 2 K, where $C_{es}$ is the electronic specific heat of the superconducting state. $C_{es}$ is obtained by subtracting the phonon specific heat $C_{ph} = \beta T^3 + \delta T^5$ from $C$(0 T) and extrapolating $C_{es}$(0 T) to 0 K. The entropy difference goes to zero, when the temperature reaches 1.8 K, as shown in the inset, which confirms the thermodynamic consistency of the fitting.

In Fig. 5(b), the difference in $C/T$ between 0 T and 0.2 T, $[C(0\ T) - C(0.2\ T)] / T$, is shown as a function of the temperature. The dashed line represents an entropy-conserving construction. The transition to the superconducting state with an entropy-conserving construction gives $\Delta C(T_C) / T_C$ = 11.8(2) mJ/mol$K^2$ and $T_C$ = 1.66 K. The ratio of the specific heat jump at $T_C$ to $\gamma_n$, $\Delta C(T_C)/\gamma_n T_C$, is calculated to be 1.37, which is almost the same as the value of the BCS theory of 1.43.[40]

The anomaly in the $C/T$-$T$ curve has a shoulder at 1.76 K. A detailed investigation of the shoulder is underway, although one plausible origin is spatial inhomogeneity of the chemical composition in the samples.

Figure 6(a) shows a magnetic field-temperature diagram deduced from the measurements of electrical resistivity, ac magnetic susceptibility, and specific heat. The open triangle, circle and rhombus symbols indicate $T_C^{\rho_on}$, $T_C^{\rho_mid}$, and $T_C^{\rho_comp}$ shown in Fig 3(b), respectively. The solid squares are $H_{C2}$ determined from χ', $H_{C2}^{\chi'}$, as shown in the inset of Fig. 4. The solid triangles are $T_C$ values determined from the specific heat data $T_C^{SH}$ using entropy-conserving construction.

Figure 6 (b) shows a reduced magnetic field $h^*$ as a function of the reduced temperature $t^*$, where $h^* = H / (-dH/dT)T_C$ and $t^* = T / T_C$. The dashed-dotted and dotted lines show $h^*$ calculated by a pair-breaking model in the clean limit and in the dirty limit,[41-43] respectively. The $h^*$-$t^*$ curve obtained from the resistivity data agrees with those from the ac magnetic susceptibility and specific heat data. The $t^*$ dependence of $h^*$ deviates from the calculated curve. We, therefore, determined $H_{C2}(0)$ to be 0.27 T from $\mu_0 H_{C2}^{\rho\text{–comp}}$. This value of $\mu_0 H_{C2}(0)$ is approximately one-tenth the Pauli limiting field $\mu_0 H_P$ (3.31 T), defined as $H_P = 18.4 T_C$ (kOe),[44, 45] which indicates an absence of Pauli limiting in $BaPt_2Sb_2$. The Ginzburg-Landau (GL) coherent length $\xi(0)$ was determined to be 35 nm using the formula $H_{C2}(0) = \Phi_0/2\pi\xi(0)^2$, where $\Phi_0$ is the flux quantum.[40]

## C. Electronic Structure

Figures 7 and 8 illustrate the band structure and the total and partial density of states (DOS) of $BaPt_2Sb_2$, respectively. The bands in the energy range from -0.2 to 0.0 Ry mainly consist of Sb-s states. The states in the energy range from 0.1 to 0.7 Ry mainly consist of Sb-p and Pt-d. Above 0.7 Ry, the bands originate from the Ba-d and Pt-d states. Since the band widths of Pt and Sb are broad, the correlation is deduced to be weak in $BaPt_2Sb_2$. Four bands cross the Fermi level $E_F$, as can be clearly seen in the Γ - Y line. The DOS at the Fermi level ($N(E_F)$) is 29.8 states/Ry/chemical formula unit or 2.58 states/eV/chemical formula unit. Figure 8(b) illustrates an expanded view of the partial DOS of $BaPt_2Sb_2$ near the Fermi level $E_F$. $N(E_F)$ mainly consists of Sb-p and Pt-d states and the contribution from the Pt-d states is larger in $N(E_F)$, especially that from the d states of Pt1 atoms (Pt atoms at the 4g site in Table I) are largest. The magnitudes of the contribution to $N(E_F)$ of Pt2-d, Sb2-p, and Sb1-p are approximately one half, one third, and one fifth of that from the Pt1-d states, respectively. It is worth noting that the Pt1 atom is the atom that forms the $PtSb_4$ tetrahedra in the $Pt_2Sb_2$ layers. These results suggest that the carriers mainly conduct in the $Pt_2Sb_2$ layers formed from the $PtSb_4$ tetrahedra.

As described above, the four bands cross the $E_F$. The bands that cross the $E_F$ in the Γ - Y line from the Γ to Y points correspond to the 37th, 38th, 39th, and 40th bands from the bottom, respectively, which results in four sheets of FSs. Figure 9 illustrates a separate sheet of the Fermi surfaces (FSs). Two of these sheets (the FSs of the 37th and 38th bands) are hole-like and the rest are electron-like. The 37th and 38th bands form 3D FSs around the Γ point. The FSs of the 39th and 40th bands have 2D FSs with cylindrical shape along the $k_z$ direction at the corners around the M point.

The shape of the FSs in $BaPt_2Sb_2$ is close to those observed in $BaPd_2Sb_2$[29] and $SrPt_2As_2$[22]--they have two 3D FS and two 2D FSs. On the other hand, the carrier types of FSs are different between $SrPt_2As_2$ and the other two antimonides; $SrPt_2As_2$ has one hole-like FS and three electron-like FSs, while $BaPt_2Sb_2$ has two hole-like FSs and two electron-like FSs, as in the case of $BaPd_2Sb_2$.

The anisotropic shapes of the FSs are expected to be reflected in anisotropic transport properties. In order to see it, we calculated the Fermi velocity $v_F^x$ and $v_F^z$ assuming the angle $\beta_{uc}$ in the unit cell to be 90° because the first two digits in the Fermi velocities are the same between the crystal structure with $\beta_{uc}$ of 90.00° and $\beta_{uc}$ of 91.227°. Table 3 lists averaged $v_F^x$, $v_F^z$ and the ratio $v_F^x/v_F^z$ of $BaPt_2Sb_2$ together with the two polymorphs of $BaPd_2Sb_2$ and $SrPt_2As_2$. The values of the ratio $v_F^x/v_F^z$ of $BaPt_2Sb_2$ and the $CaBe_2Ge_2$-type materials are clearly larger than one, which suggests that these materials have anisotropic transport properties. On the other hand, these values for the $ThCr_2Si_2$-type materials are almost one, suggesting isotropic transport properties. Thus, the existence of 2D FSs in $BaPt_2Sb_2$ and the $CaBe_2Ge_2$-type materials is reflected in anisotropic transport properties.

Table IV shows the superconducting critical temperatures $T_C$, electronic heat capacity coefficients $\gamma_n$, Debye temperatures $\theta_D$, density of states at the Fermi level $N(E_F)$, and electron-phonon coupling constants $\lambda$ of $BaPt_2Sb_2$, $SrPt_2Sb_2$,[30] $LaPd_2Sb_2$,[46] and $SrPt_2As_2$.[20,22] $LaPd_2Sb_2$ is a recently reported superconducting material ($T_C$ = 1.4 K) with the $CaBe_2Ge_2$-type structure. $\gamma_n^{exp}$ is $\gamma_n$ obtained from specific heat measurements and $\gamma_n^{cal}$ is $\gamma_n$ calculated using the equation: $\gamma_n^{cal} = (\pi^2/3)N(E_F)k_B^2$. The electron-phonon coupling constant $\lambda$ was estimated to be 0.67 for $BaPt_2Sb_2$ using the relation: $\gamma_n^{exp} = (1+\lambda)\gamma_n^{cal}$. $BaPt_2Sb_2$ is a superconducting material with moderate coupling. The value of $\gamma_n^{exp}$ of $BaPt_2Sb_2$ is comparable to those of $SrPt_2Sb_2$ and $SrPt_2Sb_2$, and is slightly larger than that of $LaPd_2Sb_2$. $\theta_D$ of $BaPt_2Sb_2$ is the smallest among the four compounds. The value of $\lambda$ is comparable to that of $SrPt_2As_2$.

## IV. CONCLUSION

This study clearly demonstrates that the 122-type antimonide, $BaPt_2Sb_2$, structurally related to the iron-based superconductors, shows superconductivity with a $T_C$ of 1.8 K. $BaPt_2Sb_2$ crystallizes in a new-type of monoclinic variant of the $CaBe_2Ge_2$-type structure. The calculation shows that it has two 3D FSs and two 2D FSs, which results in anisotropic transport properties. The d-states of Pt atoms from $PtSb_4$ tetrahedra in the $Pt_2Sb_2$ layers mainly contribute to $N(E_F)$. A comparison between experimental and calculated results indicates that $BaPt_2Sb_2$ is a superconducting material with moderate coupling.

This study also demonstrates that unknown variants of $CaBe_2Ge_2$-type structure plausibly exist in the 122-type pnictides. Since it is expected that there is room for searching for superconducting materials in the 122-type antimonides, the search of antimonides with the $CaBe_2Ge_2$-type structure or its variants is interesting, and gives us a chance to discover new superconducting materials.

Since the calculation of the electronic structure of $CaBe_2Ge_2$-type iron-free pnicitdes shows that they have intermediate character between the $ThCr_2Si_2$-type iron pnictides and the $ThCr_2Si_2$-type iron-free pnictides, further study of the 122-type antimonides with the $CaBe_2Ge_2$-type structure or its variants will be important for a systematic understanding of superconductivity in the 122 pnictides.

Acknowledgements

The authors thank M. Nishio of NIMS for the EPMA, M. Takagi of NIMS for experimental support, Y. Yamada, F. Ishikawa, N. Eguchi of Niigata University for their support at the first stage in the synthesis. This work was supported in part by the Japan Society for the Promotion of Science (JSPS) through its 'Funding Program for World-Leading Innovative R&D on Science and Technology (FIRST Program)'.

Table I. Crystal structure of $BaPt_2Sb_2$. Monoclinic. Space group: C2/m (No. 12). Lattice parameters: $a$ = 6.70156(10) Å, $b$ = 6.75246(10) Å, $c$ = 10.47440(14) Å, $\alpha_{uc}$ = 90.000°, $\beta_{uc}$ = 91.2274(9)°, $\gamma_{uc}$ = 90.000°. $Z$ = 4. $U_{iso}$=0.0057 Å$^2$ (fixed). R-factors: $R_p$ = 9.86, $R_{wp}$ = 12.0, $R_e$ = 4.57, $\chi^2$ = 6.939. The definitions of $R_p$, $R_{wp}$, $R_e$ and $\chi^2$ are described in Ref. [31]. The figures in parentheses represent the standard deviation.

| Label | Atom | Multiplicity Wyckoff position | x | y | z |
|---|---|---|---|---|---|
| Ba | Ba | 4i | 0.2549(6) | 0.0000 | 0.7531(4) |
| Pt1 | Pt | 4g | 0.0000 | 0.2823(3) | 0.0000 |
| Pt2 | Pt | 4i | 0.2472(3) | 0.0000 | 0.3791(2) |
| Sb1 | Sb | 4h | 0.00000 | 0.2505(4) | 0.50000 |
| Sb2 | Sb | 4i | 0.2276(5) | 0.00000 | 0.1206(3) |

Table 2. Interatomic distances and bond angles. The figures in parentheses represent the standard deviation.

| | | |
|---|---|---|
| Interatomic distances | | |
| Intralayer | | |
| $Pt_2Sb_2$ layer | 2.681(3) Å x 2 | 2.734(3) Å x2 |
| | Sb height from the Pt square lattice | 1.263(3) Å |
| $Sb_2Pt_2$ layer | 2.687(2) Å x2 | 2.702(2) Å x2 |
| | Pt height from the Sb square lattice | 1.266(2) Å |
| Interlayer | 2.708(7) Å | |
| Bond angles | | |
| $PtSb_4$ tetrahedron | | |
| Sb-Pt-Sb angle | | |
| diagonal ($\alpha_{bond}$) | 122.92(15)° x 2 | |
| adjacent ($\beta_{bond}$) | 91.58(16)°, 102.78(12)° x 2, 113.50(19)° | |
| $SbPt_4$ tetrahedron | | |
| Pt-Sb-Pt angle | | |
| diagonal ($\alpha_{bond}$) | 123.95(12)° x 2 | |
| adjacent ($\beta_{bond}$) | 102.47(12)°, 103.10(9)° x 2, 102.34(12)° | |

Table III. Averaged Fermi velocity $v_F^x$, $v_F^z$ and the ratio $v_F^x/v_F^z$ for $BaPt_2Sb_2$ together with those of two polymorphs ($CaBe_2Ge_2$-type and $ThCr_2Si_2$-type) of $BaPd_2Sb_2$[29] and $SrPt_2As_2$[22].

| Materials | $v_F^x$ ($10^7$cm/s) | $v_F^z$ ($10^7$cm/s) | $v_F^x/v_F^z$ |
|---|---|---|---|
| $BaPt_2Sb_2$ | 4.4 | 2.1 | 2.1 |
| The $CaBe_2Ge_2$ type | | | |
| $BaPd_2Sb_2$ | - | - | 2.9 [a] |
| $SrPt_2As_2$ | 2.47 [b] | 1.62 [b] | 1.52 [b] |
| The $ThCr_2Si_2$ type | | | |
| $BaPd_2Sb_2$ | - | - | 0.72 [a] |
| $SrPt_2As_2$ | 1.93 [b] | 1.87 [b] | 1.03 [b] |

[a] Ref. 29

[b] Ref. 22

Table IV. Superconducting critical temperatures $T_C$, electronic heat capacity coefficients $\gamma_n$, Debye temperatures $\theta_D$, calculated density of states at the Fermi level $N(E_F)$, and electron-phonon coupling constants $\lambda$ of $BaPt_2Sb_2$, $SrPt_2Sb_2$[30], $LaPd_2Sb_2$ [46] and $SrPt_2As_2$[20, 22]. $\gamma_n^{exp}$ is $\gamma_n$ obtained from specific heat measurements and $\gamma_n^{cal}$ is $\gamma_n$ estimated by calculations. The figures in parentheses represent the standard deviation.

| | $T_C$ (K) | $\gamma_n^{exp}$ (mJ/mol $K^2$) | $\theta_D$ (K) | $N(E_F)$ (states/eV cell) | $\gamma_n^{cal}$ (mJ/mol $K^2$) | $\lambda$ |
|---|---|---|---|---|---|---|
| $BaPt_2Sb_2$ | 1.82 | 8.6(2) | 146(4) | 2.19 | 5.16 | 0.67 |
| $SrPt_2Sb_2$ | 2.1[a] | 9.2(1)[a] | 183[a] | - | | - |
| $LaPd_2Sb_2$ | 1.4[b] | 6.89[b] | 210[b] | | | |
| $SrPt_2As_2$ | 5.2[c] | 9.72[c] | 211[c] | 2.55[d] | 6.01[d] | 0.62[d] |

[a] Ref. 30

[b]Ref. 46

[c] Ref. 20

[d] Ref. 22

Fig. 1. (Color online) Powder X-ray diffraction pattern of $BaPt_2Sb_2$. The red points and black line represent the observed and calculated intensities, respectively. The difference of the two intensities is shown by the blue line shifted by -5000 counts. Peak positions for $BaPt_2Sb_2$ and the impurity PtSb are labeled by green and purple vertical bars located at -1000 and -2000 counts, respectively.

Fig. 2. (Color online) (a) Crystal structure of $BaPt_2Sb_2$, (b) $PtSb_4$ and $SbPt_4$ tetrahedra that form the $Pt_2Sb_2$ and $Sb_2Pt_2$ layers, (c) view from the [010] direction, (d) view from the [111] direction, (e) deformed Pt square lattice of $BaPt_2Sb_2$ from the [001] direction, and (f) Pt square lattice of $CaBe_2Ge_2$-type $BaPt_2Sb_2$ from the [001] direction. Red broken lines show the Pt lattice. The blue solid lines show the unit cell. Large purple spheres, small white spheres, and small orange spheres represent Ba, Pt, and Sb atoms, respectively.

Fig. 3. (Color online) (a) Electrical resistivity as a function of temperature, (b) electrical resistivity at various constant temperatures as a function of magnetic field. The inset in Fig. 3 (a) is an expanded view of the resistivity at low temperatures ranging from 1.8 to 2.2 K. The symbols $H_{C2}^{\rho_on}$, $H_{C2}^{\rho_mid}$, $H_{C2}^{\rho_comp}$, represent the magnetic field at which $\rho$ starts to increase from zero resistivity, the value of $\rho$ at half of the normal resistivity value, and an extrapolation of the $\rho$-$T$ curve during the transition that reaches the normal resistivity value, respectively.

Fig. 4. (Color online) Real and imaginary parts of the ac magnetic susceptibility, $\chi'$ and $\chi''$, measured at constant temperature below 1.44 K as a function of magnetic field amplitude $\mu_0 H$. The inset shows an expanded view of $\chi'$ measured at 0.03 K. The symbol $H_{C2}^{\chi'}$ represents a magnetic field amplitude where the diamagnetic signal disappears.

Fig. 5. (Color online) (a) Specific heat divided by the temperature $C/T$ measured at various magnetic fields as a function of temperature squared $T^2$, and (b) specific heat difference between 0 T and 0.2 T divided by temperature $[C(0\ \mathrm{T}) - C(0.2\ \mathrm{T})] / T$ as a function of the temperature.

Fig. 6. (Color online) (a) Magnetic field-temperature diagram deduced from measurements of electrical resistivity, ac magnetic susceptibility, and specific heat. The open triangle, circle and rhombus symbols indicate $T_C^{\rho_on}$, $T_C^{\rho_mid}$, and $T_C^{\rho_comp}$, as shown in Fig 3(b). The solid square and triangle symbols are $H_{C2}$ determined from $\chi'$ and $T_C$ determined from the specific heat data, respectively. (b) Reduced magnetic field $h^*$ as a function of the reduced temperature $t^*$, where $h^* = H / (-dH/dT)T_C$ and $t^* = T / T_C$. The dashed-dotted and dotted lines show $h^*$ calculated by a pair-breaking model in the clean limit and in the dirty limit, respectively.

Fig. 7. Band structure of $BaPt_2Sb_2$.

Fig. 8. (Color online) (a) Total and partial density of states (DOS) and partial DOS of $BaPt_2Sb_2$. (b) Expanded view of partial DOS of $BaPt_2Sb_2$ near the Fermi level $E_F$.

Fig. 9. Fermi surfaces (FSs) of $BaPt_2Sb_2$. FSs of the 37th and 38th bands are hole-like and FSs of the 39th and 40th bands are electron-like.

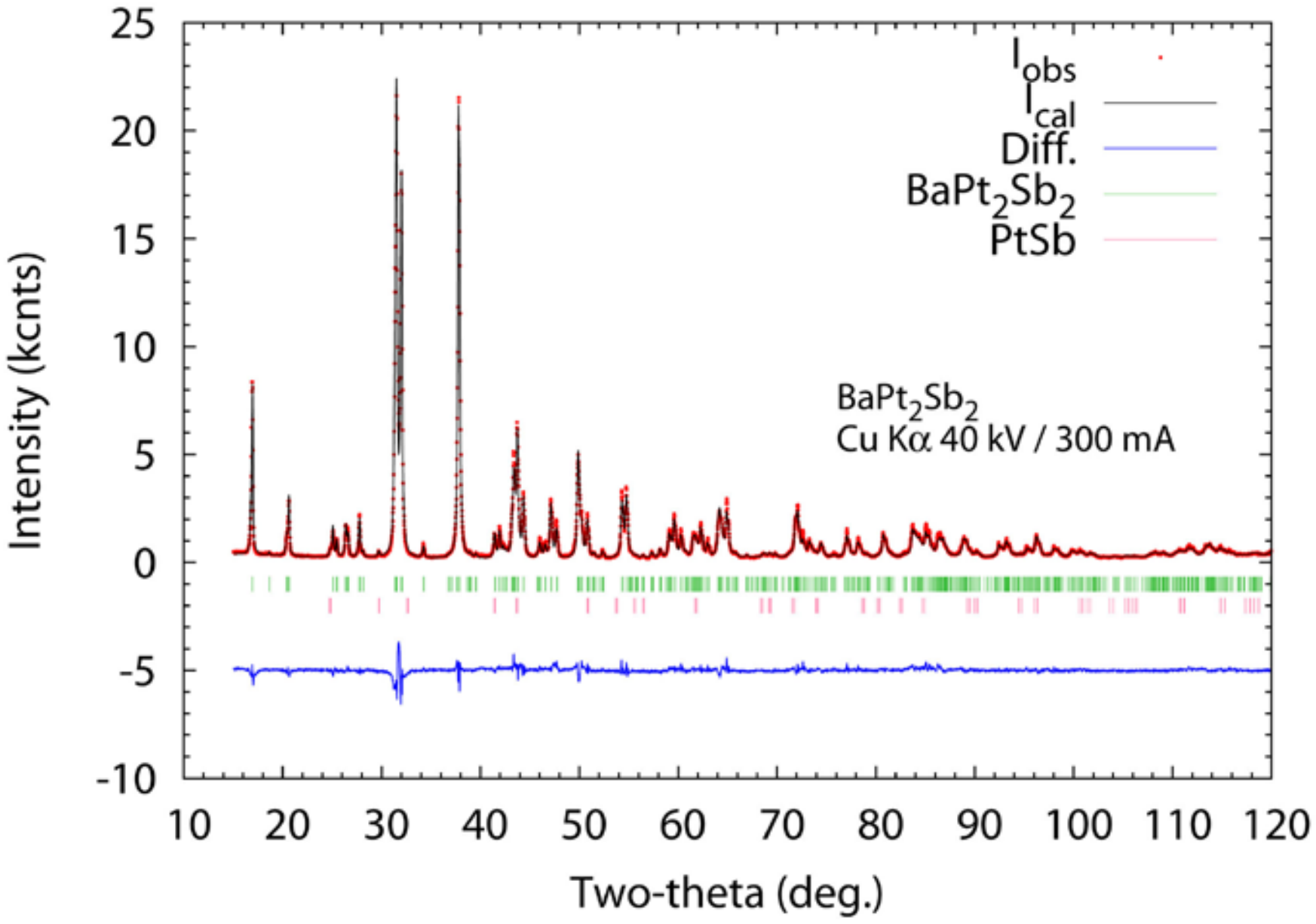

Fig. 1

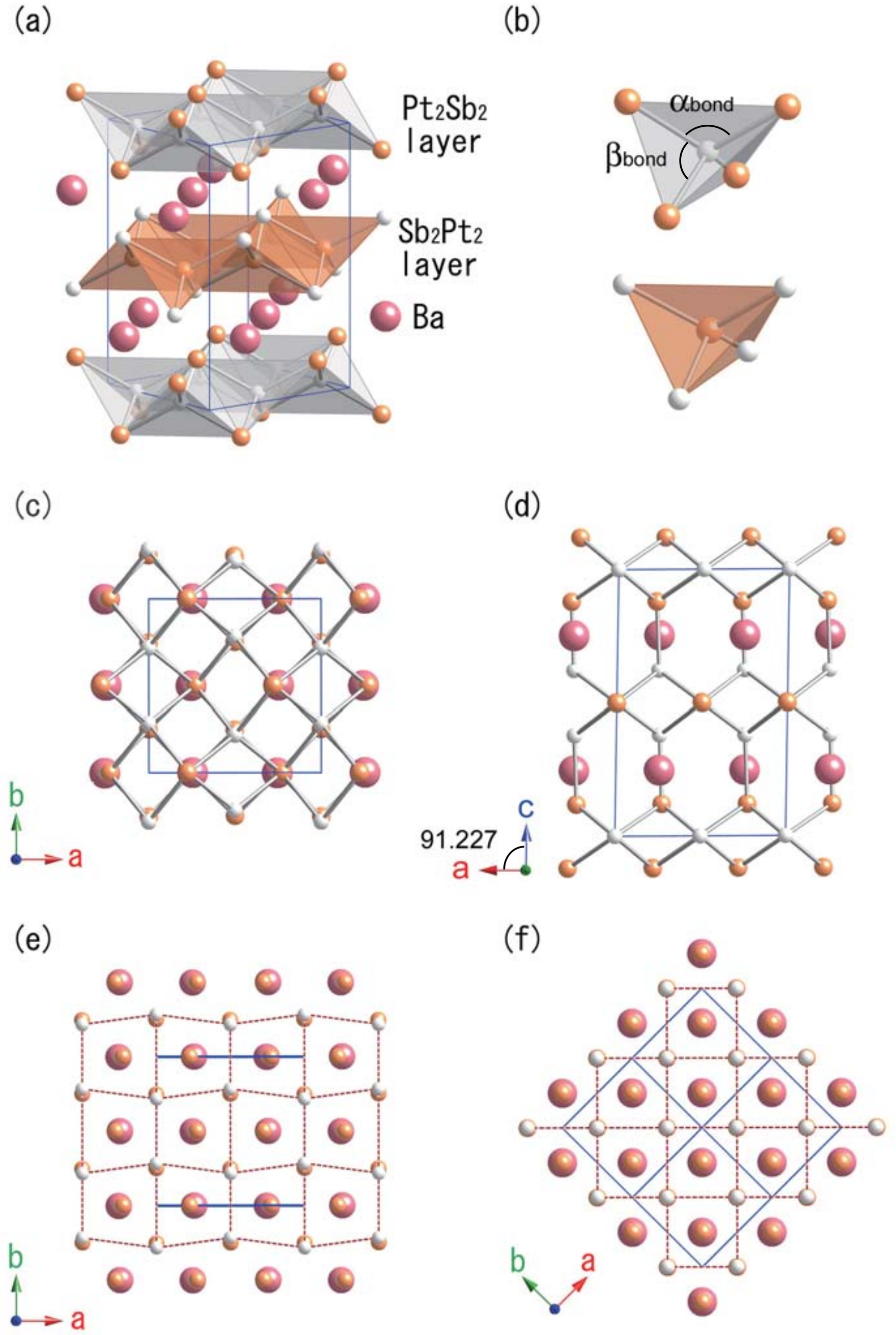

Fig. 2

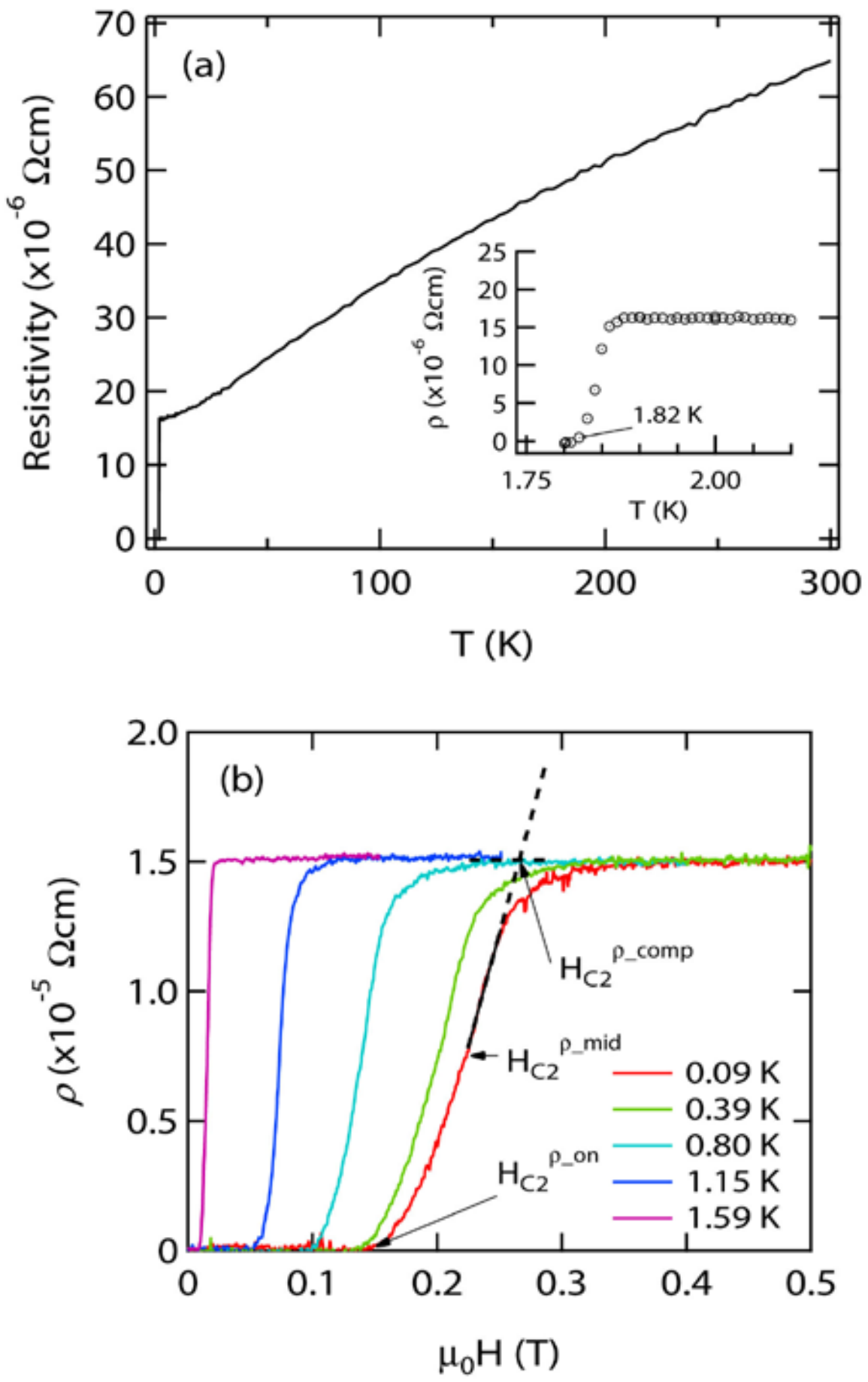

Fig. 3

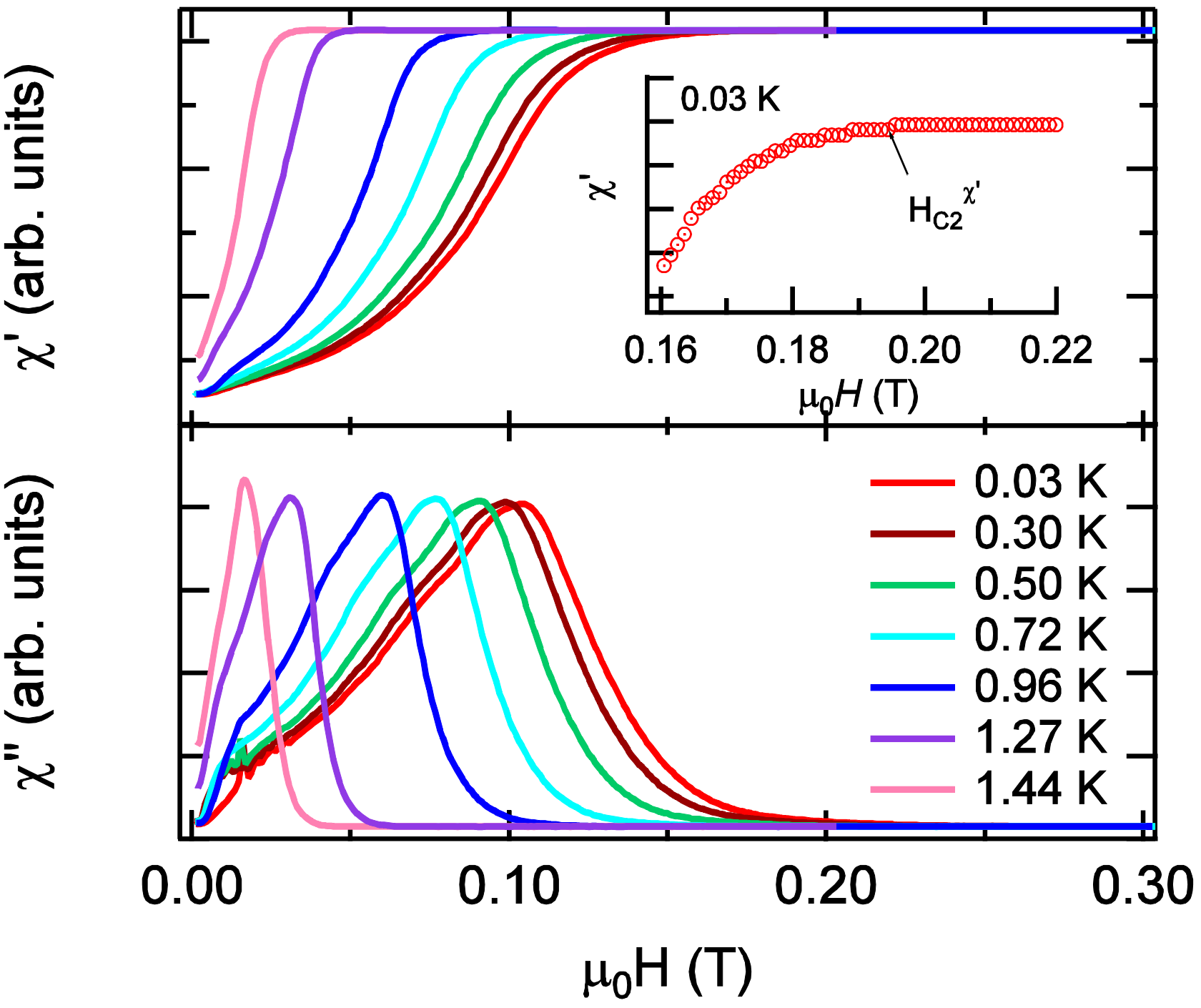

Fig. 4

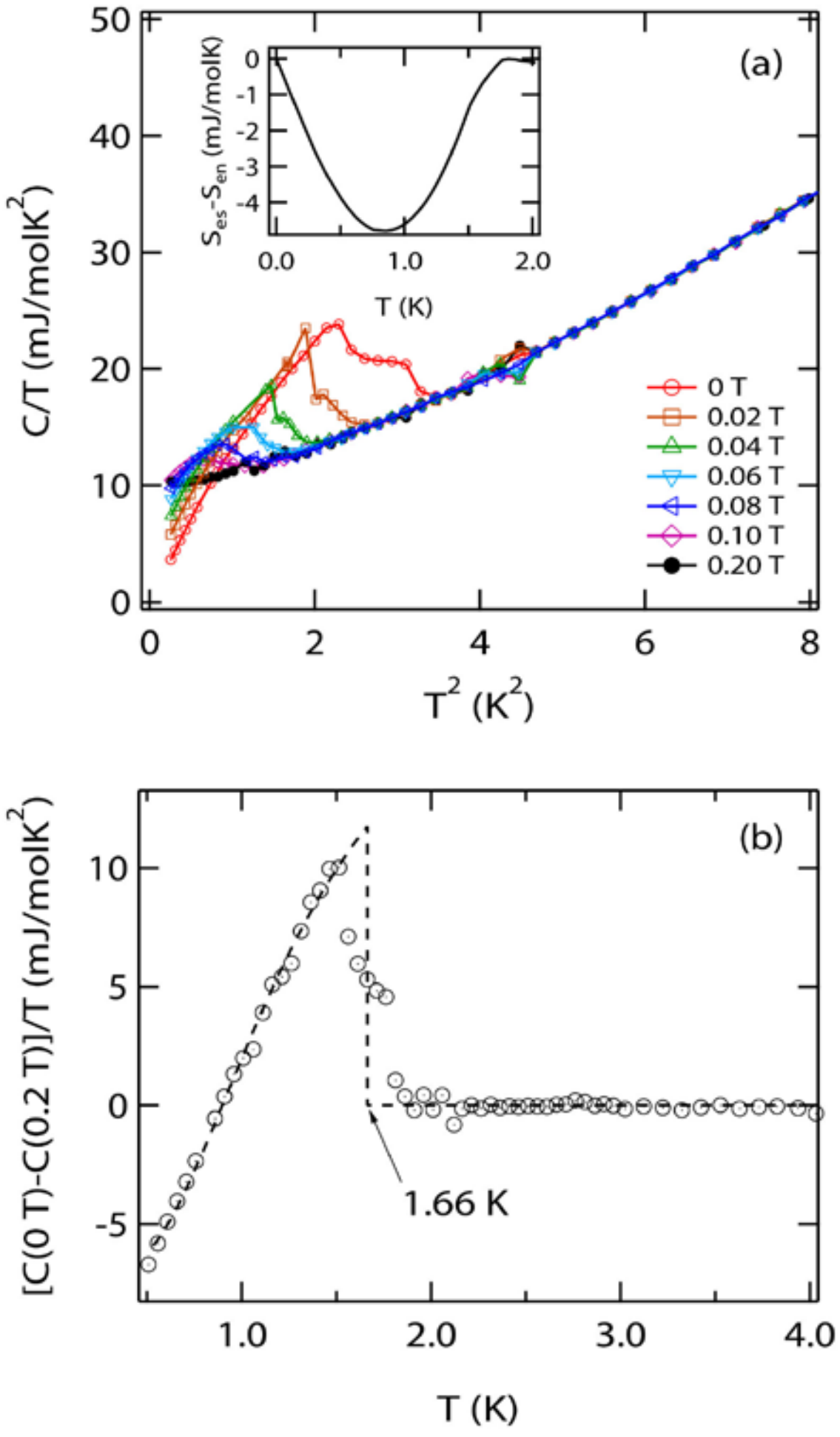

Fig. 5

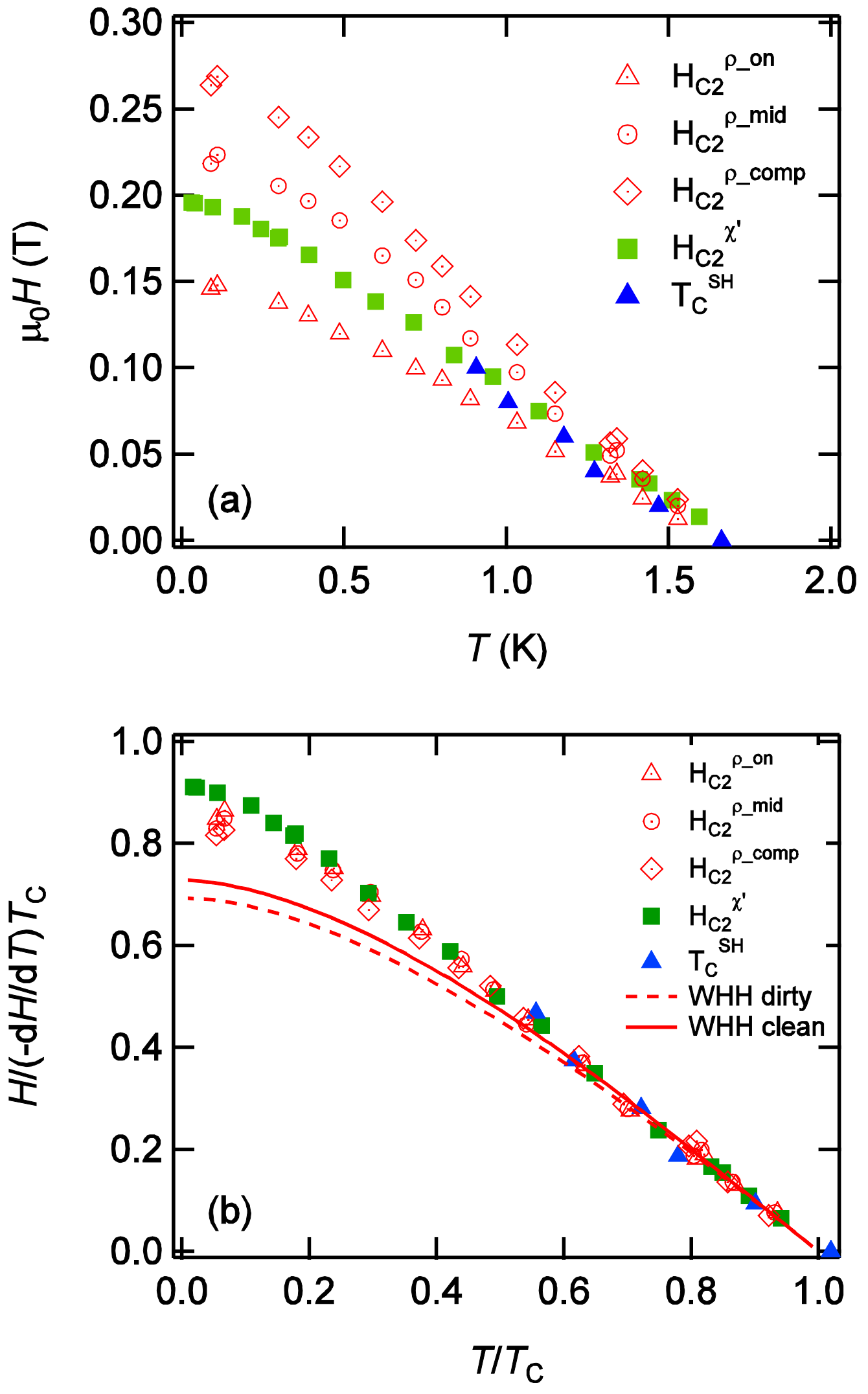

Fig. 6

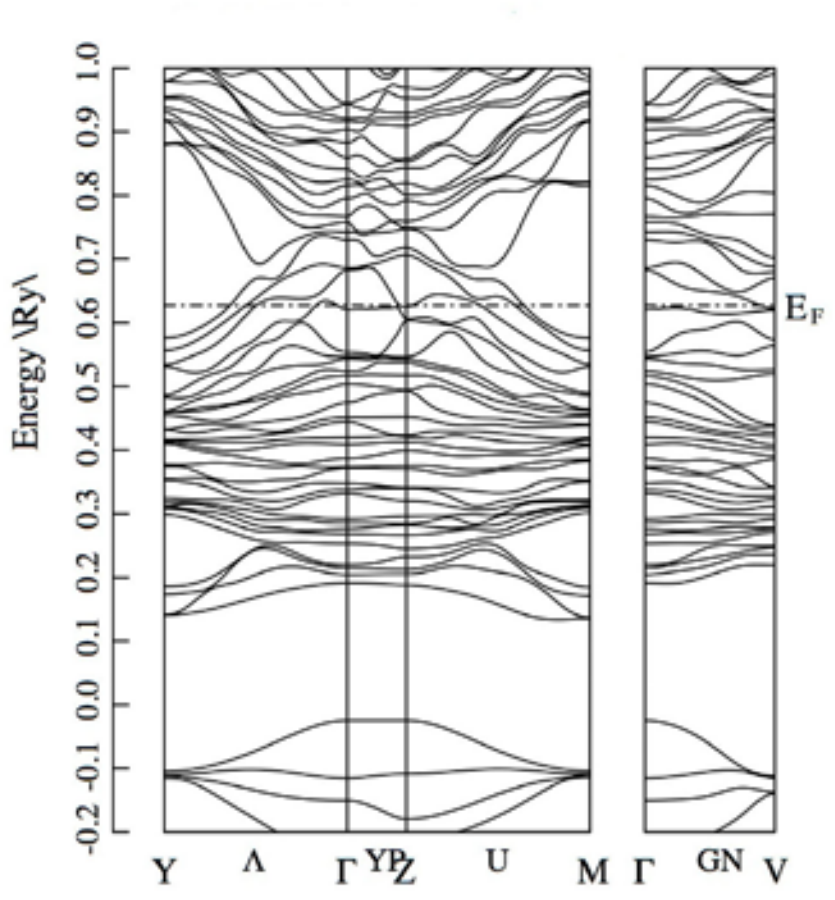

Fig. 7

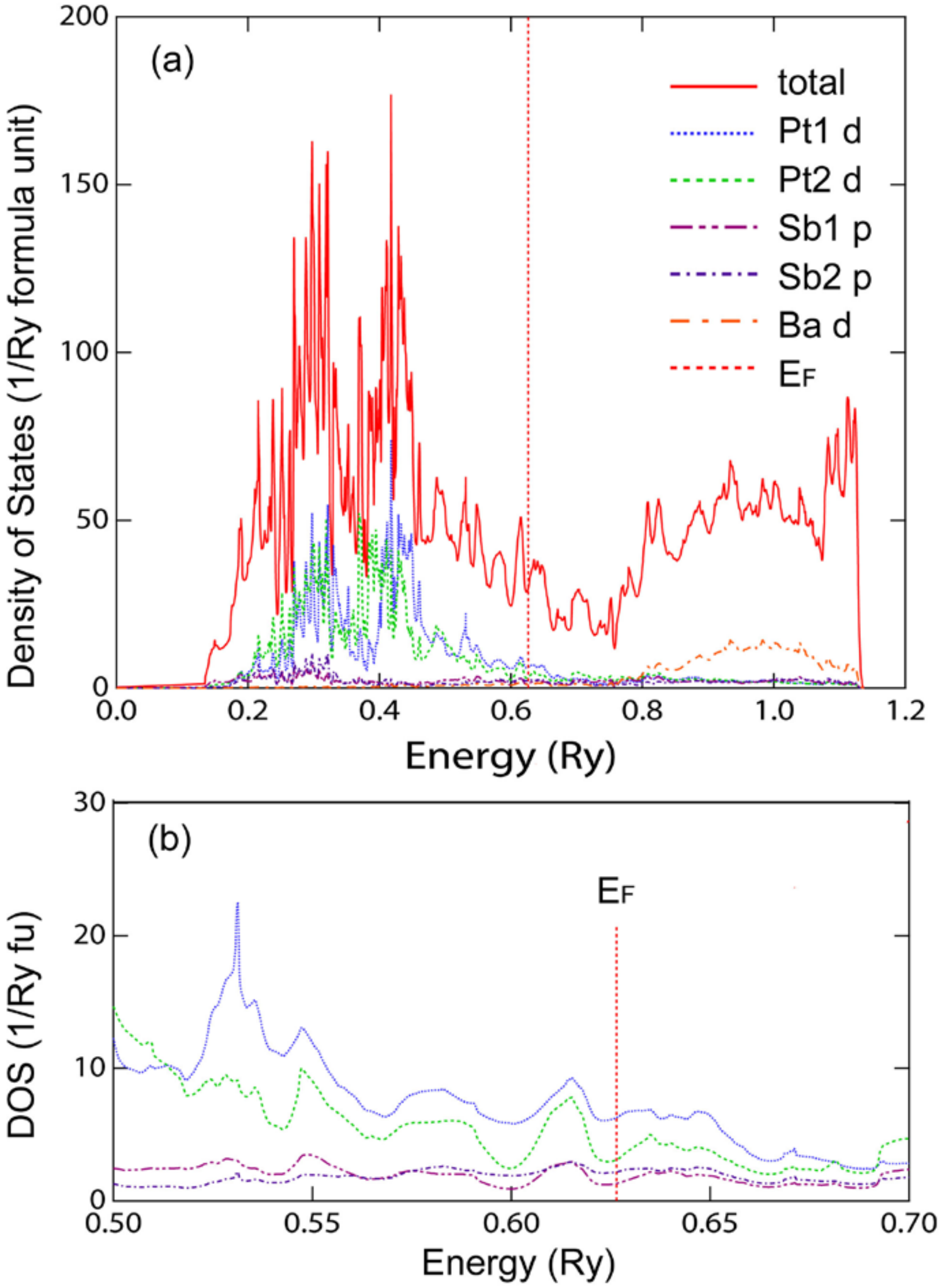

Fig. 8

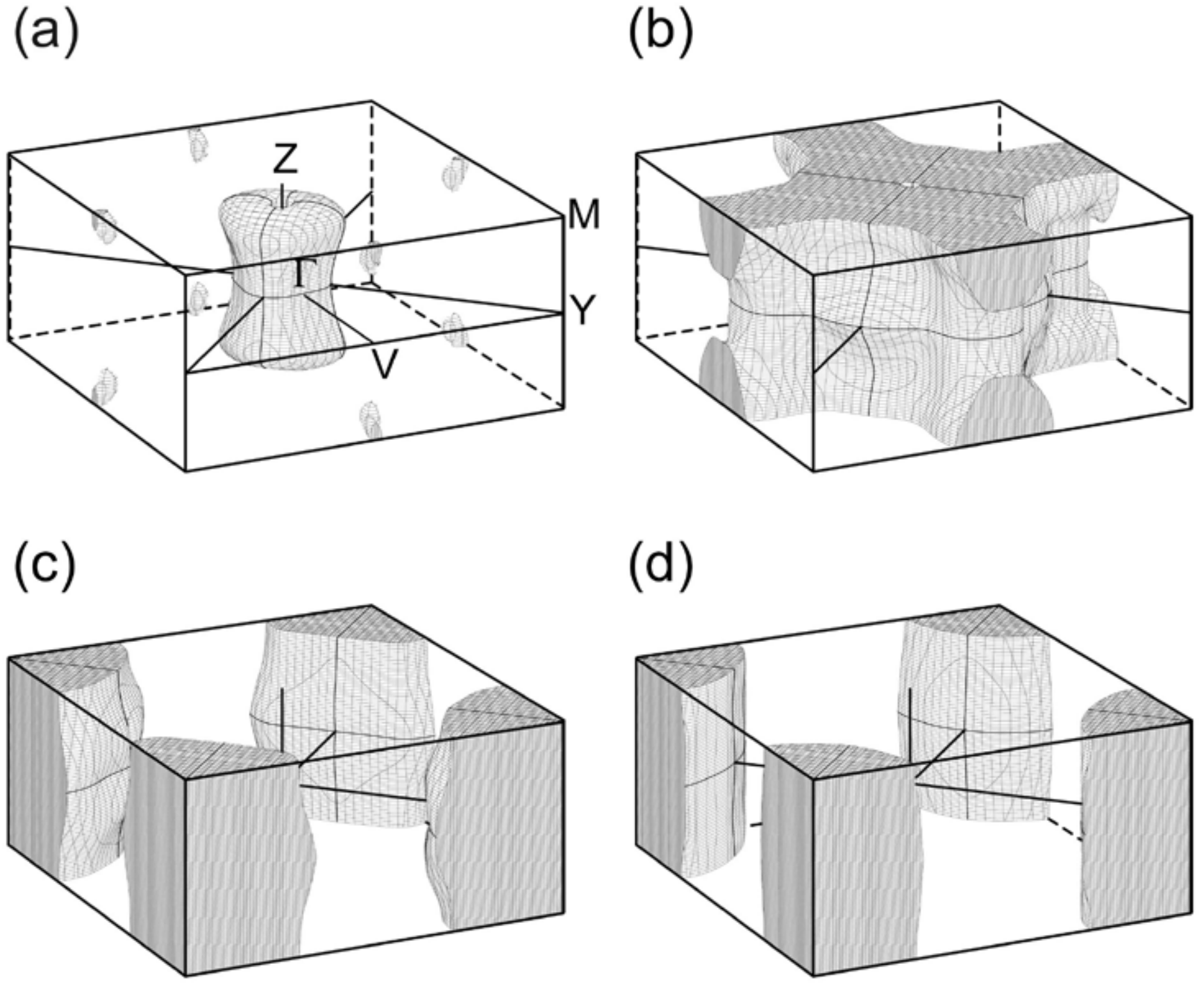

Fig. 9